\documentclass[%
 reprint,
superscriptaddress,
 amsmath,amssymb,
aps,
pra,
]{revtex4-2}

\usepackage{graphicx}
\usepackage{dcolumn}
\usepackage{bm}
\usepackage{color}

\usepackage{booktabs}
\usepackage{makecell}
\usepackage{array}
\usepackage{tabularx}

\begin{document}

\hyphenpenalty=2000
\preprint{APS/123-QED}

\title{Time-resolving the birth of photoelectrons in strong-filed ionization \\ 
with an isolated attosecond pulse}

\author{Kunlong Liu}
\email{liukunlong@gbu.edu.cn} 
\affiliation{School of Physical Sciences, Great Bay University, Dongguan 523000, China}

\author{Yidian Tian}
\email{d202280096@hust.edu.cn}
\affiliation{School of Physics and Wuhan National Laboratory for Optoelectronics, Huazhong University of Science and Technology, Wuhan 430074, China}

\author{Peng-Cheng Li}
\email{pchli@stu.edu.cn}
\affiliation{Research Center for Advanced Optics and Photoelectronics, Department of Physics, College of Science, Shantou University, Shantou, Guangdong 515063, China}

\date{\today}

\begin{abstract}
To time-resolve attosecond electronic dynamics in general photoionization processes, the technique that retrieves the phase of emitted electronic wave packets without intercepting the interactions is essential. Here, we theoretically demonstrate a scheme that uses isolated attosecond pulses (IAPs) to achieve this goal. Our approach utilizes the coherent interference between the electronic wave packets of interest and the one produced by a subsequent IAP. It is shown that the photoelectron spectral phase that has eluded direct detection so far can be fully recovered from observable photoelectron spectra without perturbing the electron-release process under investigation. By further performing a time-frequency-like analysis on the photoelectron energy spectra with the spectral phase, we reveal the birth processes of photoelectrons in time and the association between electronic energy and birth time in strong-field ionization driven by circularly polarized laser pulses. The present work explores a promising application of IAPs for ultrafast measurement and opens a viable venue for investigating electronic dynamics with quantum phase information. 
\end{abstract}

\maketitle

\newpage
When an atom is struck by a strong laser pulse, the bound electrons have the chance to escape from the core and become photoelectrons \cite{Keldysh, Faisal, Reiss}. This process, known as strong-field ionization (SFI), is a primary step for a rich set of ultrafast phenomena in attosecond physics \cite{Corkum}.
Quantum mechanically, the burst of the electronic wave packet (EWP) occurs every laser cycle during the SFI \cite{Uiberacker}, and the energy and timing information regarding the electron-release process is eventually encoded in the amplitude and phase of the photoelectron spectrum \cite{Yakovlev,Gruson,Pengju}. However, the phase of the EWP is inaccessible in routine time-independent photoelectron spectroscopy, making it difficult for us to fully time-resolve the birth of photoelectrons in SFI processes.

The instantaneous ionization rate of the SFI inside a laser cycle was studied in theory over two decades ago \cite{Yudin}. Yet, it is still challenging to capture experimentally the temporal electronic signal within a specific laser cycle. 
Although remarkable techniques such as attoclock \cite{Eckle} and photoelectron holography \cite{Huismans} have been proposed to study the subcycle SFI dynamics, neither scheme identifies the photoelectrons born in neighboring laser cycles.
Key to overcoming the present challenge is accessing the spectral phase of photoelectrons without perturbing the ionization. A landmark experimental study recently demonstrated the feasibility of reading out the phase evolution of photoelectrons after SFI using the Kapitza-Dirac effect \cite{Lin}, but it is still insufficient to reveal the association between electronic energy and birth time for the complete SFI process.

To resolve the SFI in time, we need an event at least on the attosecond time scale.  
Thanks to advances in high-harmonic generation \cite{McPherson, Ferray} and the development of X-ray free-electron lasers \cite{Ackermann, Emma, Ishikawa, Allaria, Kang, Altarelli, Milne, Zhao}, isolated attosecond pulses (IAPs) are now available in laboratories \cite{Prat, Hentschel, Sansone, Goulielmakis, ZhaoK, Zhan,Li, Gaumnitz, Yang, Wang, Takahashi, Fu, Hartmann, Duris}.
With such an ultrafast probe, a compelling question arises: is it possible to utilize the event induced by the IAP to clock the SFI \textit{without} intercepting the electron-release process?

Let us begin with the scenario where a single ionization burst (without rescattering) occurs under the driving pulse peaked at $t=0$. 
For simplicity, we focus on the freed photoelectron with the final momentum direction of $(\theta_k=\pi/2$,\ $\varphi_k=0)$, i.e.~in parallel to $\mathbf{e}_x$.
At the moment $t$ sufficiently long after the interaction, the photoelectron can be represented by (in atomic units) \cite{Pazourek, Ivanov2}
\begin{eqnarray}
\label{eq:wave0}
\psi_0(E;t) &=&a_0(E) e^{i(\phi_0+\phi_\mathrm{i})} e^{-i E[t-\tau_0(E)]}   ,
\end{eqnarray}
where $a_0(E)$ denotes the real amplitude, $\phi_\mathrm{0}$ the total $E$-independent phase offset caused by the interaction, and $\phi_\mathrm{i}$ the unknown initial phase of the system.
The term $E[t-\tau_{0}(E)]$ suggests the accumulated phase of a free EWP propagating in the continuum of energy $E$ since $\tau_{0}(E)$.
Here, we regard $\tau_{0}$ as the intrinsic birth time of the photoelectron, since $\tau_{0}$ is imprinted on the outgoing EWP through the interaction and remains unchanged in free propagation. 
So far, the fact that the EWP phase is hidden from conventional detectors has prevented us from accessing $\tau_{0}$ directly.
In this study, we will demonstrate, based on experimental observables, the feasibility of exposing the birth time of photoelectrons by applying an IAP subsequently to the ionization, and reveal the energy-time characteristics of photoelectrons in multiphoton and tunneling ionization.

When the target EWP is free, a subsequent IAP is applied to the system at $t=t_\mathrm{x}$, inducing a reference EWP represented by
$\psi_\mathrm{x}(E;t)=
a_\mathrm{x}(E) e^{i (\phi_\mathrm{x}+\phi_\mathrm{i} )} e^{-iE[t-\tau_\mathrm{x}(E)]} 
 $ with `X' denoting the quantities associated with the IAP.  
{Then, the photoelectron energy spectrum (PES) around the direction of $(\theta_k=\pi/2,\ \varphi_k=0)$ is given by }
{
\begin{eqnarray}
\label{eq:PE} 
P (E,t_\mathrm{x})&=&\left|\psi_0(E,t)+\psi_\mathrm{x}(E,t)\right|^2 \\ \nonumber
&=& P_0(E) +P_\mathrm{x}(E)   + 2 a_0  a_\mathrm{x}  \cos(\tau_\mathrm{E} E + \phi_0-\phi_{\mathrm{x}} ).
\end{eqnarray}
}Here, $\tau_\mathrm{E}=\tau_0(E)-\tau_\mathrm{x}(E)$ indicates the birth delay between two EWPs. {$P_0(E)=|\psi_0(E,t)|^2=a_0^2(E)$ and $P_\mathrm{x}(E)=|\psi_\mathrm{x}(E,t)|^2=a_\mathrm{x}^2(E)$ represent the PES} when only the driving pulse and only the IAP are applied, respectively.
Note that $\tau_0(E)$ is now encoded in the interference term. 
Based on Eq.~(\ref{eq:PE}), {a real wave packet retrieved from a set of spectra [$P(E,t_\mathrm{x})$, $P_0(E)$, and $P_\mathrm{x}(E)$] is defined as}
\begin{eqnarray}
\label{eq:fE}
W(E,t_\mathrm{x})
:&=&\frac{P(E,t_\mathrm{x})-P_0(E)-P_\mathrm{x}(E)}{2\sqrt{P_\mathrm{x}(E)}}
\\ \nonumber
&=& 
{ a_0(E)\cos(\tau_\mathrm{E} E + \phi_0-\phi_{\mathrm{x}} )}
\\ \nonumber
&\approx&
{  a_0(E)\cos(\tau_\mathrm{D} E +\phi_0-\phi_{\mathrm{x}}),}
\end{eqnarray}
where $\tau_\mathrm{D}=\tau_0(E)-t_\mathrm{x}<0$ indicates the birth delay of the target EWP with respect to the moment of $t=t_\mathrm{x}$.
In Eq.~(\ref{eq:fE}), we assumed $\tau_\mathrm{x}(E) \approx t_\mathrm{x}$ (thus, $\tau_\mathrm{E} \approx \tau_\mathrm{D}$), i.e.~an approximately instant electronic transition by the IAP compared to the ionization process by the driving pulse.
As we shall show, this assumption is basically justified for chirp-free IAPs, although the retrieved EWPs for SFI deviate slightly from the exact ones.

On the other hand, based on Eq.~(\ref{eq:wave0}), we rewrite the target EWP at $t=t_\mathrm{x}$ as
\begin{eqnarray}
\label{eq:psit0}
{\Psi_0(E,t_\mathrm{x}):= \psi_0(E;t=t_\mathrm{x})
= a_0(E)e^{i(\tau_\mathrm{D} E+\phi_0+\phi _\mathrm{i})}. \ }
\end{eqnarray}  
By comparing Eqs.~(\ref{eq:fE}) and (\ref{eq:psit0}), we find that upon the approximation, $W(E,t_\mathrm{x})$ and $\Psi_0(E,t_\mathrm{x} )$ share the same amplitude and relative phase. For sufficiently large $t_\mathrm{x}$, the phase term in Eq.~(\ref{eq:fE}) would oscillate much faster than the modulation of the amplitude. Then, the energy-dependent phase of $W(E,t_\mathrm{x})$, denoted as $\phi_W^\mathrm{H}(E,t_\mathrm{x})$, can be numerically extracted based on the Hilbert transform \cite{Cohen, Abdelhakiem}. 
Therefore, by applying an IAP to the system at a given time after ionization, one can retrieve from observables the target EWP of that given moment, just like capturing the ionizing EWP with an attosecond flash.

\begin{figure}[t]
\begin{center}
\includegraphics[width=8.6 cm]{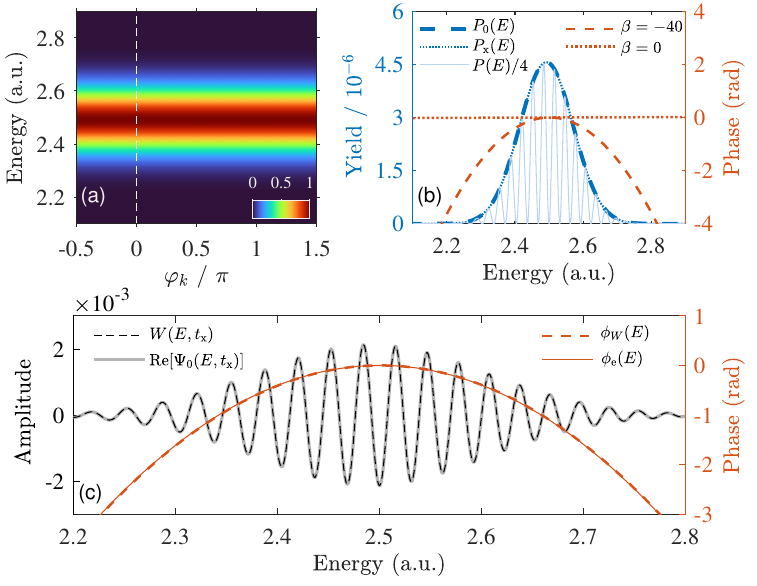}
\caption{\label{fig:xuv} The ionization scenario induced by the chirped IAP with $\mathcal{E}_0=0.01\ \mathrm{a.u.}$, $\omega_0=3\ \mathrm{a.u.}$, $\tau_\mathcal{E}=30\ \mathrm{a.u.}$, and $\beta=-40$. (a) The angle-resolved PES (normalized to the maximum, in linear color scale). The dashed line indicates $\varphi_k=0$. (b) The PES cut at $\varphi_k=0$ and the spectral phases. (c) The theoretically exact and {retrieved wave packets} and their phases. See text for details.}
\end{center}
\end{figure}

To verify the theory above, we numerically solved the time-dependent Schr\"{o}dinger equation (TDSE) for laser-driven ionization of atomic hydrogen {within the dipole approximation} \cite{FEDVR,splitlanczos,Volkov} (see
Appendix A for a brief description of the numerical method). The electric field for the laser pulse at the delay of $\Delta t$ is given by $\vec{\mathcal{E}}(t)= \mathrm{Re}[\mathcal{E}_\mathrm{c}(t-\Delta t)]\mathbf{e}_x+\mathrm{Im}[\mathcal{E}_\mathrm{c}(t-\Delta t)]\mathbf{e}_y$, where
$\mathcal{E}_\mathrm{c}(t)  =  \frac{\mathcal{E}_0}{\sqrt{ \sigma }} 
e^{-  \frac{\alpha}{\sigma} t^2 }
e^{i\left(\omega_0 t  +  \phi_\mathrm{CE}  \right)} $
with $\alpha=(8\ln 2 )/ \tau^2_\mathcal{E}$ and $\sigma=1+4i\alpha\beta$, so that the pulse spectrum remains the same when the linear chirp characterized by $\beta$ is introduced (see
Appendix A for the derivation). We set $\Delta t\equiv0$ for driving pulses and $\Delta t=t_\mathrm{x}$ for the IAP. $\mathcal{E}_0$, $\omega_0$, $\tau_\mathcal{E}$, and $\phi_\mathrm{CE}\equiv \pi/2$ determine the amplitude, central frequency, pulse duration, and carrier-envelope phase, respectively, in chirp-free ($\beta=0$) cases. 
{To avoid the complexity caused by electron rescattering and focus on direct ionization,} counterclockwise circularly polarized pulses were used throughout the present study. 

\begin{figure*}[t]
\begin{center}
\includegraphics[width=17.8 cm]{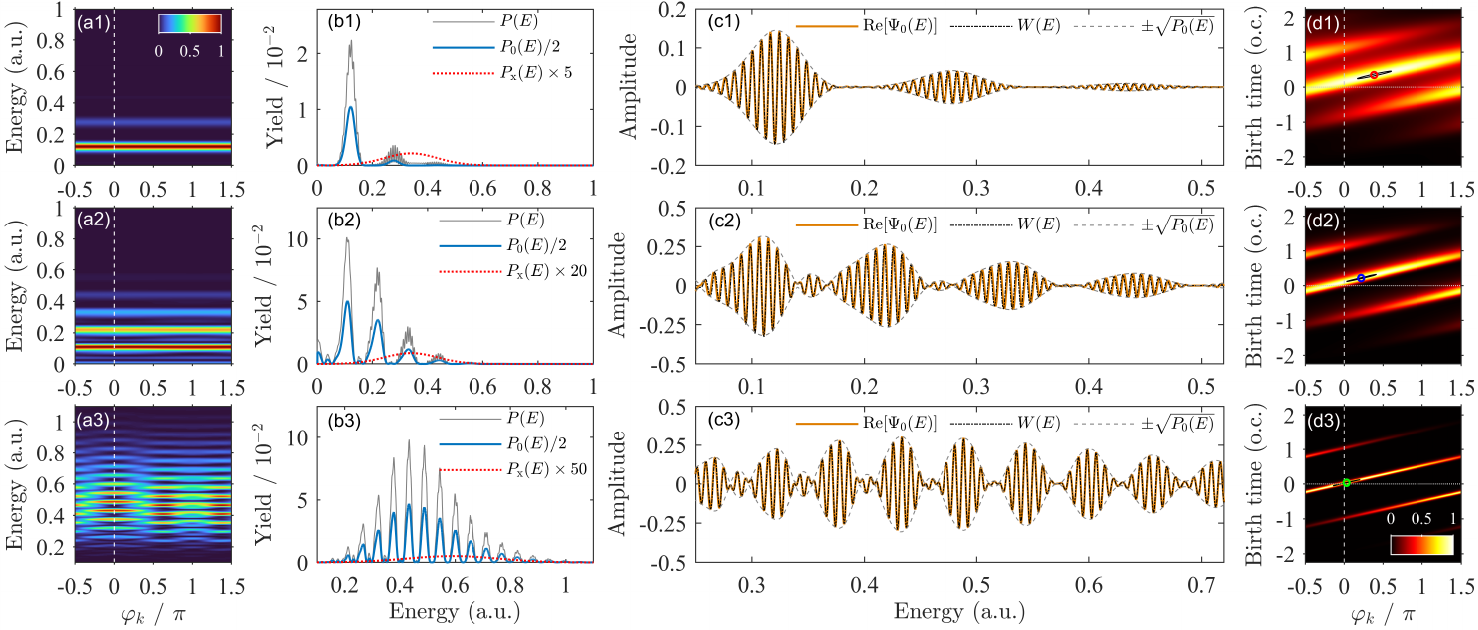}
\caption{\label{fig:sfi} The SFI scenarios driven by the circularly polarized pulses.
(a1)--(a3) The angle-resolved PES (normalized to the maximum, in linear color scale) induced by the driving pulses. (b1)--(b3) The PES cut at $\varphi_k=0$. The PES multiplied by factors are for better visibility. (c1)--(c3) The exact and {retrieved wave packets} and the corresponding amplitudes at $\varphi_k=0$.
(d1)--(d3) The angle-resolved BTDs (normalized to the maximum, in linear color scale), where the circles located at {$\varphi_k=0.383\pi,\ 0.216\pi,\ \mathrm{and}\ 0.0306\pi$} indicate the overall maxima, respectively, and the solid contours indicate $99\%$ of the maxima. The vertical and horizontal dashed lines indicate $\varphi_k=0$ and $\tau_0=0$, respectively. 
{From top to bottom, the laser parameters of the driving pulses are
$\mathcal{E}_0=0.03$, $0.045$, and $0.05\ \mathrm{a.u.}$, $\omega_0=0.16$, $0.1139$, and $0.05695\ \mathrm{a.u.}$, and $\tau_\mathcal{E} \equiv 7\ \mathrm{o.c.}$,
respectively, while those of the IAPs are $\mathcal{E}_0=0.01$, $0.01$, and $0.015\ \mathrm{a.u.}$, $\omega_0=0.88$, $0.88$, and $1.2\ \mathrm{a.u.}$, and $\tau_\mathcal{E} = 25,\ 25,\ \mathrm{and}\ 15\  \mathrm{a.u.}$, respectively. Note that the same IAP is applied for the first and second rows.}}
\end{center}
\end{figure*}

In the numerical experiments for the particular scenario discussed above, we chose a linearly chirped IAP to trigger the ionization and its chirp-free version to induce the reference EWP. The angle-resolved PES under the chirped IAP is shown in Fig.~\ref{fig:xuv}(a). 
The PES cut at $\varphi_k=0$ [$P_0(E)$ and $P_\mathrm{x}(E)$] for the chirped and chirp-free IAPs (as individual driving pulses), as well as their spectral phases, are shown in Fig.~\ref{fig:xuv}(b), together with $P(E)$ for applying these two pulses with a delay of $t_\mathrm{x}=200\ \mathrm{a.u.}$
The PES for the chirped and chirp-free IAPs are identical, while the spectral phases tell the different underlying events.
So far, the theoretical spectral phase is calculated by
$\phi_\mathrm{e} (E):=\mathrm{arg}[\Psi_0 (E,t_\mathrm{x})] +t_\mathrm{x}E$, with the exact {wave packet} $\Psi_0 (E,t_\mathrm{x})$ obtained from the TDSE. 
According to Eq.~(\ref{eq:psit0}), the straight PES phase shown in Fig.~\ref{fig:xuv}(b) suggests a constant birth time of the photoelectron for the chirp-free IAP, demonstrating the justification of the assumption made in Eq.~(\ref{eq:fE}), while the curved phase shows that the birth time varies with the energy under the chirped IAP.

Then, with the three PES shown in Fig.~\ref{fig:xuv}(b), we reconstruct the real {wave packet} according to Eq.~(\ref{eq:fE}) and the result is shown in Fig.~\ref{fig:xuv}(c), together with the retrieved spectral phase given by $\phi_W (E):= \phi_W^{\mathrm{H} }(E,t_\mathrm{x})  +t_\mathrm{x}E$.
For comparison, we also show $\phi_\mathrm{e} (E)$ and the real part of $\Psi_0(E,t_\mathrm{x})$, whose global phase has been shifted so that its phase at the PES maximum aligns with that of $W(E,t_\mathrm{x})$. It is clear that the {retrieved wave packet} is consistent with the exact one upon a global phase shift.

Next, we extend our scheme to general SFI scenarios,
where the ionization burst takes place every laser cycle.
Similarly to Eq.~(\ref{eq:psit0}), the multiple ionizing EWPs propagating until $t=t_\mathrm{x}$ are represented by 
\begin{eqnarray}
\label{eq:waven}
\Psi_0'(E,t_\mathrm{x}) 
= {\sum_{n}  a_{0,n}(E)e^{i(\tau_{\mathrm{D},n} E + \phi_{0,n}+\phi_{\mathrm{i}})}\ \ \  }
\end{eqnarray}
with $\tau_{\mathrm{D},n}= \tau_{0,n}(E)-t_\mathrm{x}$, where $n$ indicates the $n$th ionization burst.
Following the similar derivation and assumption for Eq.~(\ref{eq:fE}), we obtain 
\begin{eqnarray}
\label{eq:fE2}
W'(E,t_\mathrm{x})
&:=&\frac{P'(E,t_\mathrm{x})-P'_0(E)-P_\mathrm{x}(E)}{2\sqrt{P_\mathrm{x}(E)}}
\\ \nonumber 
&\approx& { \sum_n   a_{0,n}(E) \cos(\tau_{\mathrm{D},n} E +\phi_{0,n} - \phi_{\mathrm{x}}),}
\end{eqnarray}
with {$P'(E)=|\psi'_0(E,t)+\psi_\mathrm{x}(E,t)|^2$ and $P'_0(E)=|\psi'_0(E)|^2$} indicating the PES for applying two sequential pulses and only the driving pulse, respectively. 
Equations (\ref{eq:waven}) and (\ref{eq:fE2}) show that recovery of photoelectron {wave packets} for SFI scenarios would also work, provided that the IAP produces sufficient ionization yields to effectively interfere with the target EWP in the energy domain. In addition, the EWPs of different momentum directions can be retrieved following the same procedure.

For demonstration, in the first three columns of Fig.~\ref{fig:sfi} we present the angle-resolved PES, the cut PES, and the {wave packets}, respectively, for three representative sets of numerical experiments, the same as those in Fig.~\ref{fig:xuv} but with different IAPs applied at $t_\mathrm{x}=800\ \mathrm{a.u.}$ The Keldysh parameters for the SFI, defined as $\gamma=\omega_0\sqrt{2I_\mathrm{p}}/\mathcal{E}_0$ with the ionization potential $I_\mathrm{p}$, are $\gamma_1=5.33 ,\ \gamma_2=2.53 ,\  \mathrm{and}\ \gamma_3=1.14 $, respectively, {ranging from the multiphoton to nonadiabatic tunneling regimes}. 
As shown in Figs.~\ref{fig:sfi}(c1)--\ref{fig:sfi}(c3), the {retrieved wave packets} are in good agreement with the exact ones. 
Slight deviations are found in Figs.~\ref{fig:sfi}(c1) and~\ref{fig:sfi}(c2), as the assumption of constant birth time for the IAPs is less accurate for slower electrons ({$E\lesssim 0.2\ \mathrm{a.u.}$}) due to the Coulomb effect \cite{Pazourek}.

\begin{figure*}[t]
\begin{center}
\includegraphics[width=17.8 cm]{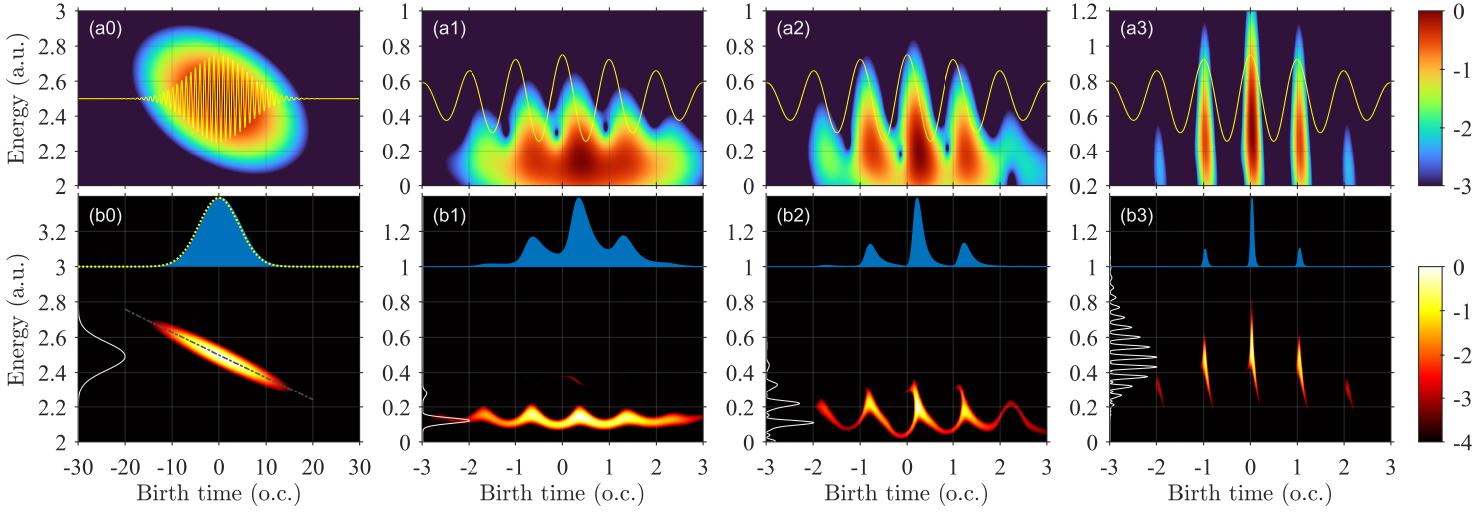}
\caption{\label{fig:sst} The energy-time representations of photoelectrons in different ionization scenarios (see text), based on GT (top row) and SST (bottom row).  The solid curves in the upper panels indicate the $y$ components of the circularly polarized driving pulses. 
The shadows and solid curves in the lower panels indicate the profiles of the corresponding BTDs and PES at selected momentum directions, respectively. In (b0), the dotted curve shows the profile of the instantaneous pulse intensity and the dash-dotted line the instantaneous pulse frequency minus $I_\mathrm{p}$. All distributions are normalized to the corresponding maxima and displayed in logarithmic color scale. {The laser parameters for the columns from left to right are the same as those in Figs.~\ref{fig:xuv}(a), \ref{fig:sfi}(a1), \ref{fig:sfi}(a2), and \ref{fig:sfi}(a3), respectively.}
}
\end{center}
\end{figure*}

Once the {wave packet} is retrieved from the observable PES (or calculated theoretically), the underlying timing information can be revealed as follows.
According to Eq.~(\ref{eq:waven}),  $\tau_{\mathrm{D},n}$ are essentially the oscillatory frequencies of the components of $\Psi_0'(E,t_\mathrm{x})$. 
Thus, by means of the Fourier transform, we obtain the birth-time distribution (BTD) of photoelectrons via ${B}(\tau_0)=|\hat{\Psi}_\mathrm{b}(\tau_0)|^2$ with 
\begin{eqnarray}
\label{eq:btd}
\hat{\Psi}_\mathrm{b}(\tau_0) = \frac{1}{\sqrt{2\pi}}
 \int_0^\infty \Psi_0'(E,t_\mathrm{x}) e^{-i (\tau_{0}-t_\mathrm{x})E}dE\ ,
\end{eqnarray}
where $\tau_{\mathrm{D},n}$ has been replaced by $\tau_{0}-t_\mathrm{x}$ in the exponent. 
For precision, we adopt the exact {wave packets} for further analysis.
Nevertheless, the dynamic information revealed from $W'(E,t_\mathrm{x})$ are qualitatively consistent with those of $\Psi'_0(E,t_\mathrm{x})$, {as we shall show and discuss before conclusion}.

In Figs.~\ref{fig:sfi}(d1)--\ref{fig:sfi}(d3), the angle-resolved BTDs are presented for the corresponding SFI scenarios, revealing several features of the underlying dynamics. 
First, the subcycle BTD peak at a given angle means that the photoelectrons emitted to a specific direction are born
with time-distributed birth probabilities that peak at a certain moment of a laser cycle. 
Secondly, the trend that the BTD peaks drift linearly with angle is observed, which is consistent with the SFI mechanism in rotating fields \cite{Eckle}. Third, the overall BTD maxima marked by the circles are found beyond $({\varphi_k=0},\ t=0)$, indicating that the most probable birth moment of the
photoelectron is beyond the instant of the field maximum. This birth delay appears to approach zero as $\gamma$ decreases. Such delay in SFI was predicted and discussed in 2011 by Ivanov \cite{Ivanov2}.
Finally, broader stripes are found in the angle-resolved BTDs for larger $\gamma$, indicating that the birth uncertainty of the photoelectron is greater for higher nonadiabaticity. This feature is in agreement with the theory in \cite{Yudin}.

To gain further insight into the association between photoelectron kinetic energy and birth time in ionization processes, we calculated the energy-time representation (ETR) for the photoelectrons of a given momentum direction. In detail, we performed a time-frequency-like analysis on the {wave packet} $\Psi'_0(E,t_\mathrm{x})$ based on the Gabor transform (GT) \cite{Gabor} and employed the synchrosqueezing transform (SST) \cite{Sheu1, Sheu2,LiP} to sharpen the ETR resolution  (see Appendix B for the methods of time-frequency analysis). 
The results based on GT and SST are shown in Fig.~\ref{fig:sst}, where the $y$ components of the driving fields are depicted accordingly in the upper row and the profiles of the corresponding BTDs and PES are displayed in the lower row.

The ETR distributions shown in Figs.~\ref{fig:sst}(a0) and \ref{fig:sst}(b0) are for the EWP of Fig.~\ref{fig:xuv}(c), which is induced by the chirped IAP. The inclined stripe with a negative slope indicates that lower-energy photoelectrons are born at relatively later moments. In particular, the sharpened stripe in Fig.~\ref{fig:sst}(b0) aligns with the dash-dotted line given by the instantaneous frequency of the chirped IAP minus $I_\mathrm{p}$.
Meanwhile, the BTD appears to agree with the profile of $|\mathcal{\vec{E}}(t)|^2$ [dotted curve in Fig.~\ref{fig:sst}(b0)].
It intuitively shows that in single-photon ionization the ionization rate and the final energy of the photoelectron are in linear relations with the instantaneous intensity and instantaneous frequency of the IAP, respectively. In addition, no subcycle structure of the ETR is observed, 
which shows that the electronic motion hardly follows the ultrafast oscillating field.

For SFI scenarios driving by rotating fields, while the one-to-one correspondence between the emission angle and the ionization time is commonly assumed in the attoclock setup \cite{Eckle}, here
the ETR distributions shown in the last three columns of Fig.~\ref{fig:sst} [corresponding to the EWPs at the angles of the BTD maxima marked in Figs.~\ref{fig:sfi}(d1)--\ref{fig:sfi}(d3)] reveal the energy-time characteristics of the photoelectron at a given angle.
In general, when the Keldysh parameter increases from $\gamma_3$ [Fig.~\ref{fig:sst}(a3)] to $\gamma_1$ [Fig.~\ref{fig:sst}(a1)], we can see the trend that the subcycle stripes of the ETR (based on GT) gradually become tilted and broadened, resulting in wider and wider BTD peaks. 
Then, we turn to the ETR based on SST for more details. In Fig.~\ref{fig:sst}(b3) for $\gamma_3$, the stripes are almost vertically concentrated at some critical moments separated by one optical cycle.
It suggests that in the tunneling regime ({$\gamma \lesssim 1 $}), the photoelectrons in a given direction correspond mainly to a certain birth moment in each cycle.  
In contrast, inclined stripes dominate in the ETR in Fig.~\ref{fig:sst}(b2), demonstrating that the energy-time characteristic changes significantly when $\gamma$ enters the multiphoton regime ($\gamma \gg 1 $). For $\gamma_2$, the photoelectron of relatively lower energy corresponds to a later birth within each cycle.  
Furthermore, in Fig.~\ref{fig:sst}(b1) for $\gamma_1$, the dominant stripe becomes a continuous wave shape. In this case, for the energy around the first PES peak, 
the photoelectrons in the given direction could be associated with more than one birth moment within an optical cycle. 
This feature is likely attributed to the chaotic electronic motion in the fast-shaking potential well \cite{Ivanov}.

\begin{figure} 
  \centering
  \includegraphics[width=8.6 cm]{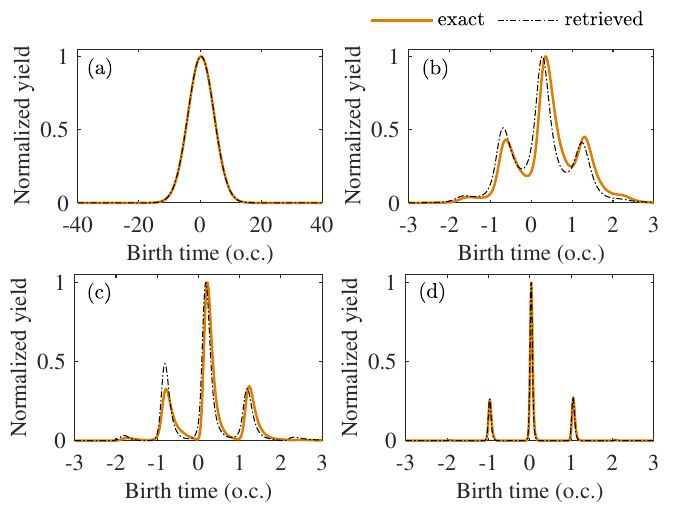}
  \caption{Comparison of the birth-time distributions obtained from the exact and {retrieved wave packets} for four ionization scenarios. {Panels (a)--(d) correspond to the scenarios that are discussed in the first to the last columns of Fig.~\ref{fig:sst}, respectively.}}
  \label{fig:BTD}
\end{figure}

\begin{figure*} 
  \centering
  \includegraphics[width=17.8 cm]{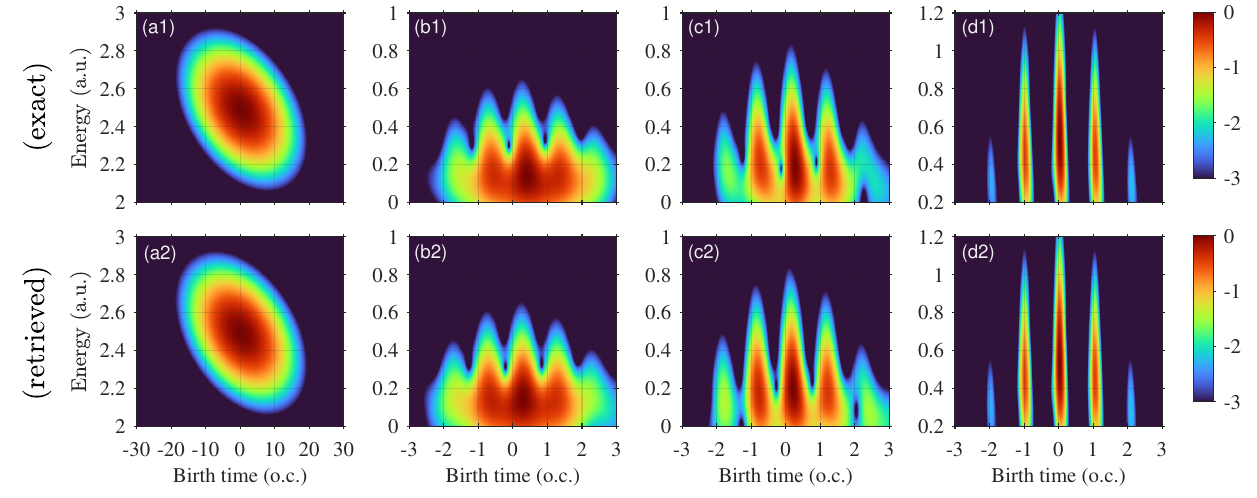}
  \caption{{Comparison of the ETR distributions based on the Gabor transform of the exact (top row) and retrieved (bottom row) wave packets.
  The four ionization scenarios from left to right correspond to those discussed in the first to the lase columns of Fig.~\ref{fig:sst}, respectively.}}
  \label{fig:GTc}
\end{figure*}

\begin{figure*}  
  \centering
  \includegraphics[width=17.8 cm]{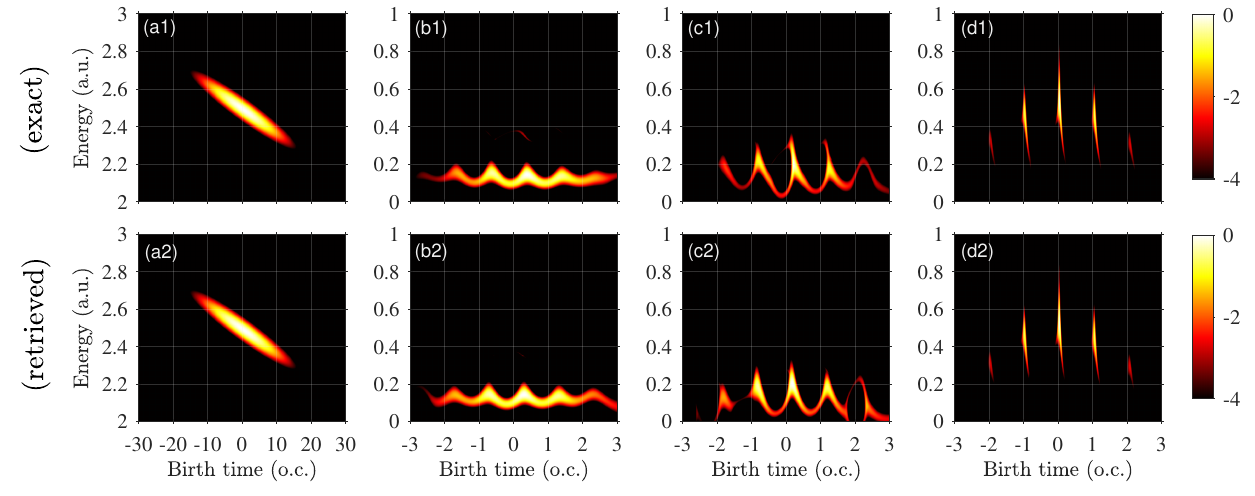}
  \caption{{Comparison of the ETR distributions based on the synchrosqueezing transform of the exact (top row) and retrieved (bottom row) wave packets.
  The four ionization scenarios from left to right correspond to those discussed in the first to the lase columns of Fig.~\ref{fig:sst}, respectively.}}
  \label{fig:SSTc}
\end{figure*}

Finally, we compare the time-frequency analysis of the exact and {retrieved wave packets} for four ionization scenarios that are discussed in Fig.~\ref{fig:sst}. The comparison for the birth-time distributions is shown in Fig.~\ref{fig:BTD}, while that for the ETR distributions based on GT and SST are shown in Fig.~\ref{fig:GTc} and Fig.~\ref{fig:SSTc}, respectively.
The momentum directions of the corresponding {wave packets} are chosen at $\theta_k=\pi/2$ and {$\varphi_k=0,\ 0.383\pi,\ 0.216\pi,\ \mathrm{and}\ 0.0306\pi$} for four scenarios, respectively.

As shown in Figs.~\ref{fig:xuv}(c) for the scenario of single-photon ionization, the {retrieved wave packets} agree very well with the exact ones. Thus, it is expected that the corresponding BTDs [see Fig.~\ref{fig:BTD}(a)], as well as the ETR distributions [see the first columns of Fig.~\ref{fig:GTc} and Fig.~\ref{fig:SSTc}], are in good agreement between the exact and {retrieved wave packets}.
For the multiphoton ionization scenarios, as shown in Figs.~\ref{fig:sfi}(c1) and \ref{fig:sfi}(c2), slight deviations can be seen between the exact and {retrieved wave packets}. Such deviations lead to small time shifts and slight profile changes of the BTDs, as observed in Figs.~\ref{fig:BTD}(b) and \ref{fig:BTD}(c), respectively. Nevertheless, the overall distributions are qualitatively in agreement between the exact and recovered results. Furthermore, by comparing the corresponding ETR distributions shown in the second and third columns of Figs.~\ref{fig:GTc} and \ref{fig:SSTc}, we can see that the ETRs based on the exact and {retrieved wave packets} are still qualitatively in agreement with each other.
For tunneling ionization, where the kinetic energies of the photoelectrons are averagely higher, the {retrieved wave packet} generally agrees with the exact one, as shown in \ref{fig:sfi}(c3). In this case, the corresponding BTDs and ETR distributions do not show much difference between the exact and retrieved wave packets [see Fig.~\ref{fig:BTD}(d) and the last columns of Fig.~\ref{fig:GTc} and Fig.~\ref{fig:SSTc}].

{Note that in the present scheme, we assumed phase-locked instant ionization by the IAP. However, as shown in the second and third columns of Fig.~\ref{fig:SSTc}, for the photoelectrons with kinetic energy lower than approximately $0.2\ \mathrm{a.u.}$, we can see the deviation of the ETR distributions between the exact and retrieved results.}
To reduce the deviation, one may need to obtain the fine energy-time structure of the IAP-generated photoelectrons via the attosecond streaking measurement in experiment \cite{Pazourek} or from theoretical calculations, and then develop the formula for a more precise reconstruction. Nevertheless, the scheme proposed in the present work provides a simple and efficient way to qualitatively reveal the information about the electronic birth time in the ionization processes of interest.

In conclusion, we have demonstrated a measurement scheme for recovering the phase information of the angle-resolved photoelectron spectrum, utilizing the coherent interference with an IAP-generated EWP. 
The principle of the method is simple and of generality for studying laser-induced ionization, as it neither depends on specific forms of the driving field nor perturbs the interaction under investigation. 
With the spectral phase being measurable, we analyzed and revealed the associations between the kinetic energy and the birth time of the photoelectrons produced in multiphoton and tunneling ionization {driven by circularly polarized pulses}, without relying on (semi)classical models. 
{In principle, the retrieval of the photoelectron spectral phase is also feasible for the ionization driven by linearly polarized pulses or other forms of driving fields that could lead to the electron rescattering, but identifying the time information about the birth and rescattering processes from the spectral phase will require further theoretical derivations.}
We anticipate that our study, combined with progress in generating shorter and stronger attosecond pulses,
will open prospects for attosecond-time-resolved observations of general electronic dynamics in molecules, nanostructures, and surfaces.

\section*{Acknowledgments}

This work is supported by National Natural Science Foundation of China (Grants No.~12174133 and No.~12434010), Natural Science Foundation of Guangdong Province (Grant No.~2025A1515011278), Department of Education of Guangdong Province (Grant No.~2024ZDZX1020), and Li Ka Shing Foundation STU-GTIIT Joint Research Grants (Grant No.~2024LKSFG02).
The computational resources are supported by SongShan Lake HPC Center (SSL-HPC) in Great Bay University.

\appendix
\section{Time-dependent Schr\"{o}dinger equation}
We numerically solved the time-dependent Schrödinger equation for the laser-driven ionization of atomic hydrogen within the dipole approximation. In the velocity gauge, the propagation of the wave function is governed by (atomic units are used):

\begin{eqnarray}
\label{eq:propagation}
i\frac{\partial \psi (\mathbf{r}, t) }{\partial t}=
\left[\frac{1}{2}\mathbf{p}^2+\mathbf{A}(t)\cdot \mathbf{p}+V(\mathbf{r}) \right]\psi (\mathbf{r}, t),
\end{eqnarray}
where  $V(\mathbf{r})=-\frac{1}{\mathbf{r}}$ is the Coulomb potential of hydrogen.

In the present study, the ionization is driven by circularly polarized laser pulses. The vector potential $\mathbf{A}(t)$ is defined via the electric field $\vec{\mathcal{E}}(t)$ as
\begin{eqnarray}
\label{eq:field1}
\mathbf{A}(t)=-\int_{-\infty}^{t} \vec{\mathcal{E}}(t')dt' \ .
\end{eqnarray}
The electric field of a laser pulse at the delay of $\Delta t$ is given by 
$\vec{\mathcal{E}}(t)= \mathrm{Re}[\mathcal{E}_\mathrm{c}(t-\Delta t)]\mathbf{e}_x+\mathrm{Im}[\mathcal{E}_\mathrm{c}(t-\Delta t)]\mathbf{e}_y$, where
\begin{eqnarray}
\label{eq:Ee} \mathcal{E}_\mathrm{c}(t)  =  \frac{\mathcal{E}_0 } {\sqrt{ \sigma }}
\exp {\left(- \frac{\alpha}{\sigma}t^2\right) }
\exp {\left[i\left(\omega_0 t  +  \phi_\mathrm{CE}  \right)\right]} 
\end{eqnarray}
with $\alpha=(8\ln 2 )/ \tau^2_\mathcal{E}$ and $\sigma=1+4i\alpha\beta$.
Note that the expression of the electric field above is obtained via the inverse Fourier transform of the complex frequency-domain pulse field $\hat{\mathcal{E}}(\omega)$, i.e.
\begin{eqnarray}
\label{eq:iEe} 
\mathcal{E}_\mathrm{c}(t)  =   \mathcal{F}^{-1}[\hat{\mathcal{E}}(\omega)]
\end{eqnarray}
with
\begin{eqnarray}
\label{eq:Ew} 
\hat{\mathcal{E}}(\omega)  &:=   &
  \sqrt{\frac{\pi}{\alpha}} \mathcal{E}_0 
\exp\left[
-\frac{(\omega-\omega_0 )^2}{4\alpha}\right] \times \\ \nonumber
 && \exp \left[-i\beta(\omega-\omega_0 )^2+i\phi_\mathrm{CE}
\right]   ,
\end{eqnarray}
so that the spectrum of the laser pulse remains unchanged when introducing the linear chirp characterized by $\beta$. We set $\Delta t\equiv0$ for the driving pulse and $\Delta t=t_\mathrm{x}$ for the delayed isolated attosecond pulse. $\mathcal{E}_0$, $\omega_0$, $\tau_\mathcal{E}$, and $\phi_\mathrm{CE}\equiv \pi/2$ determine the amplitude, central frequency, pulse duration, and carrier-envelope phase, respectively, in chirpless ($\beta=0$) cases. 
Counterclockwise circularly polarized pulses were used throughout the present study.

We propagate the time-dependent wave function $\psi (\mathbf{r}, t)$ in the spherical coordinate. The wave function $\psi (\mathbf{r}, t)$ is expanded by spherical harmonics $Y(l,m)$ as
\begin{eqnarray}
\label{eq:spherical}
\psi (\mathbf{r}, t)=\sum_{l,m}\frac{R_{l,m}(r,t)}{r}Y(l,m).
\end{eqnarray}
$R_{l,m}(r,t)$ is the radial part of the wave function, which is discretized by the finite-element discrete variable representation (FE-DVR) method \cite{FEDVR}. The angular quantum number $l$ and the magnetic quantum number $m$ are chosen up to {80}. The time-dependent wave function is propagated by the split-Lanczos method \cite{splitlanczos} with a time step of {$\delta t=0.01$ a.~u.} 
The maximum of the radial coordinate is up to 200 a.~u. The initial wave function for the ground state of hydrogen is obtained by imaginary-time propagation. In real-time propagation, the wave function is divided into the inner and outgoing parts with an absorption function, which is defined as $V_{abs}=1-[1+\exp(-(r-r_0)/\Delta r)]^{-1}$, with $r_0=120$ a.~u.~and $\Delta r=2$ a.~u. 
The inner part $\psi_{in} (\mathbf{r}, t)=\psi (\mathbf{r}, t)V_{abs}$ is propagated under the full Hamiltonian. The outer part $\psi_{out} (\mathbf{r}, t)=\psi (\mathbf{r}, t)-\psi_{in} (\mathbf{r}, t)$ is propagated by the Coulomb-Volkov propagator \cite{Volkov}. 
Eventually, at the moment $t=t_\mathrm{f}$ sufficiently long after the interaction, the ionizing wave packets are obtained by projecting the final wave function onto the scattering states \cite{splitlanczos}. 
\begin{eqnarray}
\label{eq:projection}
\psi_0(\mathbf{p};t=t_\mathrm{f})=\langle {\psi_p(\mathbf{r})}|{\psi (\mathbf{r};t=t_\mathrm{f})}\rangle,
\end{eqnarray}
where $\psi_p(\mathbf{r})$ is the normalized scattering state for hydrogen. 
{Then, the momentum-space wave function $\psi_0(\mathbf{p})$ is transformed into the angle-resolved energy-space wave function $\psi_0(E,\theta_k,\varphi_k)$ via
\begin{eqnarray}
\label{eq:ptoE}
\psi_0(E,\theta_k,\varphi_k)= (2E)^{1/4}\psi_0(p=\sqrt{2E},\theta_k,\varphi_k),
\end{eqnarray} 
so that the normalization of the wave function is preserved, i.e.
\begin{eqnarray}
\label{eq:proba}
\int\left|\psi_0(E,\theta_k,\varphi_k)\right|^2dE= \int\left|\psi_0(p,\theta_k,\varphi_k)\right|^2p^2dp
\end{eqnarray} 
with $E=p^2/2$ and $dE=pdp$.
The angle-resolved photoelectron energy spectra for $\theta_k=\pi/2$ (the polarization plane of the laser pulse) are then given by
\begin{eqnarray}
\label{eq:PES}
P(E,\theta_k=\pi/2,\varphi_k)= \left|\psi_0(E,\theta_k=\pi/2,\varphi_k)\right|^2. 
\end{eqnarray}
}

The convergence of our calculations has been confirmed by changing the maximum of the angular quantum number $l$ and magnetic quantum number $m$.

\section{Time-frequency analysis} 
Time-frequency representations provide a powerful tool for analyzing the frequency information underlying the time series signals. Here, in analogy to the time-frequency analysis of time series signals, we calculate the energy-time representations (ETRs) for the `energy series' signal, i.e.~the electronic wave packet {$\Psi_0(E,t_\mathrm{f}):= \psi_0(E;t=t_\mathrm{f})$} in the energy domain, to extract the birth time information of the photoelectron in given ranges of the kinetic energy. In the first step, we calculate the ETRs based on the Gabor transform (GT) \cite{Gabor}, which is given by
\begin{eqnarray}
\label{eq:Gabor}
\hat{g}(E,\tau )=\int^\infty_{-\infty}  
\Psi_0(E',t_\mathrm{f}) \eta(E'-E) \times\\ \nonumber
 \exp[-i\tau  (E'-E)]dE',
\end{eqnarray}
where $E$ denotes the kinetic energy. Here, the physical meaning of $\tau$ is the birth-time delay with respect to the specific moment $t=t_\mathrm{f}$ (see the main text for the discussion). 
The energy-window function $\eta(E)$ in Eq.~(\ref{eq:Gabor}) is defined as 
\begin{eqnarray}
\label{eq:Window}
\eta(E)=\frac{1}{w\sqrt{2\pi}}\exp\left({-\frac{E^2}{2w^2}}\right), 
\end{eqnarray}
where $w$ is the parameter for adjusting the width of the energy window and, in the present calculations, we set $w=0.12\ \mathrm{a.u.}$ to balance the resolutions between the energy and time domains.

Then, we employ the synchrosqueezing transform (SST) \cite{Sheu1} on the GT to sharpen the ETR resolution. Note that the SST has been shown to address the intrinsic blurring in the linear type time-frequency methods and that it is an elegant way to reveal the detailed features of quantum dynamics \cite{Sheu2,LiP}. Briefly, the SST is defined as
\begin{eqnarray}
\label{eq:SST}
\hat{s}(E,\tau)=\int \hat{g}(E,\tau')\frac{1}{\alpha_0}h\left(\frac{|\tau-\xi(E,\tau')|}{\alpha_0}\right) d\tau'
\end{eqnarray}
where $h(x)=\exp(-x^2)/\sqrt{\pi}$, $\alpha_0>0$ is a controllable smoothing parameter for the resolution, and $\xi(E,\tau')$ is the reallocation rule function defined as \cite{Sheu1}
\begin{eqnarray}
\label{eq:WF}
\xi(E,\tau')=\left\{
\begin{aligned}  &-\frac{i\partial_E \hat{g}(E,\tau')}{\hat{g}(E,\tau')} & \mathrm{when} \;\hat{g}(E,\tau')\neq 0, \\ &\;\;\;\;\;\;\;\;\;\;\;\;\infty & \mathrm{when} \;\hat{g}(E,\tau')=0.   \\\end{aligned}\right.
\end{eqnarray}
In our calculation, we choose $\alpha_0=1.5$ to balance the resolution and visibility of the results. The technical details of SST, including the accuracy and limitation, can be found in \cite{Sheu1} and the references therein.

Finally, the ETR distributions based on GT and SST are given by $G(E,\tau)=|\hat{g}(E,\tau )|^2$ and $S(E,\tau)=|\hat{s}(E,\tau )|^2$, respectively. Given that physically $\tau=\tau_0-t_\mathrm{f}$, we can eventually obtain the ETR distribution as a function of the kinetic energy $E$ and the birth time $\tau_0$ via translating $G(E,\tau)$ and $S(E,\tau)$ by $t_\mathrm{f}$ in the time dimension.


\begin{thebibliography}{99}

\bibitem{Keldysh}
L. V. Keldysh, Ionization in the field of a strong electronmagnetic wave, Sov. Phys. JETP \textbf{20}, 1945 (1964).

\bibitem{Faisal}
F. H. M. Faisal, Multiple absorption of laser photons by atoms, J. Phys. B: Atom. Mol. Phys, \textbf{6}, L89 (1973).

\bibitem{Reiss}
H. R. Reiss, Effect of an intense electromagnetic field on a weakly bound system, Phys. Rev. A \textbf{22}, 1786 (1980).

\bibitem{Corkum}
P. B. Corkum, and F. Krausz, Attosecond science, Nat. Phys. \textbf{3}, 381 (2007).


\bibitem{Uiberacker}
M. Uiberacker \textit{et al.}, Attosecond real-time observation of electron tunnelling in atoms, Nature, \textbf{446}, 627 (2007).

\bibitem{Yakovlev}
V. S. Yakovlev, F. Bammer, and A. Scrinzi, Attosecond streaking measurements, J. Mod. Opt. \textbf{52}, 395 (2005).

\bibitem{Gruson}
V. Gruson \textit{et al.}, Attosecond dynamics through
a Fano resonance: monitoring the birth of a photoelectron, Science \textbf{354}, 734 (2016).

\bibitem{Pengju}
P. Zhang, H. Liang, M. Han, J. Trester, J. Ji, J. M. Rost, and H. J. W\"{o}rner, Resolving the phase of Fano resonance wave packets with photoelectron frequency-resolved optical gating, Nat. Photonics \textbf{19}, 847 (2025).

\bibitem{Yudin}
G. L. Yudin and M. Y. Ivanov, Nonadiabatic tunnel ionization: Looking inside a Laser cycle, Phys. Rev. A \textbf{64}, 013409 (2001).


\bibitem{Ivanov}
M. Y. Ivanov, M. Spanner, and O. Smirnova, Anatomy of strong field ionization, J. Mod. Opt. \textbf{52}, 165 (2005).

\bibitem{Eckle}
P. Eckle, M. Smolarski, P. Schlup, J. Biegert, A. Staudte, M. Sch\"{o}ffler, H. G. Muller, R. D\"{o}rner, and U. Keller, Attosecond angular streaking, Nat. Phys. \textbf{4}, 565 (2008).

\bibitem{Huismans}
Y. Huismans \textit{et al.}, Time-resolved holography with photoelectrons, Science \textbf{331}, 61 (2010).

\bibitem{Lin}
K. Lin, S. Eckart, H. Liang, A. Hartung, S. Jacob, Q. Ji, L. Ph. H. Schmidt, M. S. Sch\"{o}ffler, T. Jahnke, M. Kunitski, and R. D\"{o}rner, Ultrafast Kapitza-Dirac effect, Science \textbf{383}, 1467 (2024).

\bibitem{McPherson}
A. McPherson, G. Gibson, H. Jara, U. Johann, T. S. Luk, I. A. McIntyre, K. Boyer, and C. K. Rhodes, Studies of multiphoton production of vacuum-ultraviolet radiation in the rare gases, J. Opt. Soc. Am. B \textbf{4}, 595 (1987).

\bibitem{Ferray}
M. Ferray, A. L'Huillier, X. F. Li, L. A. Lompre, G. Mainfray and C. Manus, Multiple-harmonic conversion of 1064 nm radiation in rare gases, J. Phys. B: Atom. Mol. Phys, \textbf{21}, L31, (1988).

\bibitem{Ackermann}
W. Ackermann \textit{et al.}, Operation of a free-electron laser from the extreme ultraviolet to the water window, Nat. Photonics \textbf{1}, 336 (2007).

\bibitem{Emma}
P. Emma \textit{et al.}, First lasing and operation of an \aa{ngstrom}-wavelength free-electron laser, Nat. Photonics \textbf{4}, 641 (2010).

\bibitem{Ishikawa}
T. Ishikawa \textit{et al.}, A compact X-ray free-electron laser emitting in the sub-\aa{ngstr\"{o}m} region, Nat. Photonics \textbf{6}, 540 (2012).

\bibitem{Allaria}
E. Allaria \textit{et al.}, Two-stage seeded soft-X-ray free-electron laser, Nat. Photonics \textbf{7}, 913 (2013).

\bibitem{Kang}
H.-S. Kang \textit{et al.}, Hard X-ray free-electron laser with femtosecond-scale timing jitter, Hard X-ray free-electron laser with femtosecond-scale timing jitter, Nat. Photonics \textbf{11}, 708 (2017).

\bibitem{Altarelli}
M. Altarelli, The European X-ray free-electron laser facility in Hamburg, Nucl. Instrum. Meth. B, \textbf{269}, 2845 (2011).

\bibitem{Milne}
C. J. Milne \textit{et al.},  SwissFEL: The Swiss X-ray Free Electron Laser, Appl. Sci. \textbf{7}, 720 (2017).

\bibitem{Zhao}
Z. Zhao, D. Wang, Q. Gu, L. Yin, G. Fang, M. Gu, Y. Leng, Q. Zhou, B. Liu, C. Tang, W. Huang, Z. Liu, and H. Jiang, SXFEL: A Soft X-ray Free Electron Laser in China, Synchrotron Radiat. News, \textbf{30}, 29 (2017).

\bibitem{Prat}
E. Prat \textit{et al.}, A compact and cost-effective hard X-ray free-electron laser driven by a high-brightness and low-energy electron beam, Nat. Photonics \textbf{14}, 748 (2020).

\bibitem{Hentschel}
M. Hentschel, R. Kienberger, Ch. Spielmann, G. A. Reider, N. Milosevic, T. Brabec, P. Corkum, U. Heinzmann, M. Drescher, and F. Krausz, Attosecond metrology, Nature \textbf{414}, 509 (2001).

\bibitem{Sansone}
G. Sansone \textit{et al.}, Isolated single-cycle attosecond pulses, Science \textbf{314}, 443 (2006).

\bibitem{Goulielmakis}
E. Goulielmakis \textit{et al.}, Single-cycle nonlinear optics, Science \textbf{320}, 1614 (2008).

\bibitem{ZhaoK}
K. Zhao, Q. Zhang, M. Chini, Y. Wu, X. Wang, and Z. Chang, Tailoring a 67 attosecond pulse through advantageous phase-mismatch, Opt. Lett. \textbf{37}, 3891 (2012).

\bibitem{Zhan}
M.-J. Zhan \textit{et al.}, Generation and measurement of isolated 160-attosecond XUV laser pulses at 82 eV, Chinese Phys. Lett. \textbf{30}, 093201 (2013).

\bibitem{Li}
J. Li, X. Ren, Y. Yin, K. Zhao, A. Chew, Y. Cheng, E. Cunningham, Y. Wang, S. Hu, Y. Wu, M. Chini, and Z. Chang, 53-attosecond X-ray pulses reach the carbon K-edge, Nat. Commun. \textbf{8}, 186 (2017).

\bibitem{Gaumnitz}
T. Gaumnitz, A. Jain, Y. Pertot, M. Huppert, I. Jordan, F. Ardana-Lamas, and H. J. W\"{o}rner, Streaking of 43-attosecond soft-X-ray pulses generated by a passively CEP-stable mid-infrared driver, Opt. Express \textbf{25}, 27506 (2017).

\bibitem{Yang}
Z. Yang, W. Cao, Y. Mo, H. Xu, K. Mi, P. Lan, Q. Zhang, and P. Lu, All-optical attosecond time domain interferometry, Natl. Sci. Rev. \textbf{8}, nwaa211 (2021).

\bibitem{Wang}
X. Wang, F. Xiao, J. Wang, L. Wang, B. Zhang, J. Liu, J. Zhao, and Z. Zhao, Ultrashort isolated attosecond pulse generation with 750-nm free-carrier envelope phase near-infrared pulses, Ultrafast Sci. \textbf{4}, 0080 (2024).


\bibitem{Takahashi}
E. J. Takahashi, P. Lan, O. D. M\"{u}cke, Y. Nabekawa, and K. Midorikawa, Attosecond nonlinear optics using gigawatt-scale isolated attosecond pulses, Nat. Commun. \textbf{4}, 2691 (2013).

\bibitem{Fu}
Y. Fu, K. Nishimura, R. Shao, A. Suda, K. Midorikawa, P. Lan, and E. J. Takahashi, High efficiency ultrafast water-window harmonic generation for single-shot soft X-ray spectroscopy, Commun. Phys. \textbf{3}, 92 (2020).


\bibitem{Hartmann}
N. Hartmann \textit{et al.}, Attosecond time-energy structure of X-ray free-electron laser pulses, Nat. Photonics \textbf{12}, 215 (2018).

\bibitem{Duris}
J. Duris \textit{et al.}, Tunable isolated attosecond X-ray pulses with gigawatt peak power from a free-electron laser, Nat. Photonics \textbf{14}, 30 (2020).

\bibitem{Pazourek}
R. Pazourek, S. Nagele, and J. Burgd\"{o}rfer, Attosecond chronoscopy of photoemission, Rev. Mod. Phys. \textbf{87}, 765 (2015).

\bibitem{Ivanov2}
{I. A. Ivanov, Time delay in strong-field photoionization of a hydrogen atom, Phys. Rev. A \textbf{83}, 023421 (2011).}

\bibitem{Cohen}
L. Cohen, P. Loughlin, and D. Vakman, On an ambiguity in the definition of the amplitude and phase of a signal, Signal Processing \textbf{79}, 301 (1999).

\bibitem{Abdelhakiem}
A. Matsuki, H. Kori, and R. Kobayashi, An extended Hilbert transform method for reconstructing the phase from an oscillatory signal, Sci. Rep. \textbf{13}, 3535 (2023).

\bibitem{FEDVR}
T. N. Rescigno and C. W. McCurdy, Numerical grid methods for quantum-mechanical scattering problems, Phys. Rev. A \textbf{62}, 032706 (2000).

\bibitem{splitlanczos}
W. Jiang and X. Tian, Efficient Split-Lanczos propagator for strong-field ionization of atoms, Opt. Express \textbf{25}, 26832 (2017).

\bibitem{Volkov}
D. G. Arbó, J. E. Miraglia, and M. S. Gravielle, Coulomb-Volkov approximation for near-threshold ionization by short laser pulses, Phys. Rev. A \textbf{77}, 013401 (2008).

\bibitem{Gabor}
D. Gabor, Theory of communication, J. Comm. Eng. \textbf{93}, 429 (1946).

\bibitem{Sheu1}
Y.-L. Sheu, L.-Y. Hsu, H. Wu, P.-C. Li, S.-I Chu, A new time-frequency method to reveal quantum dynamics of atomic hydrogen in intense laser pulses: Synchrosqueezing transform, AIP Adv. \textbf{4}, 22 (2014).

\bibitem{Sheu2}
Y.-L. Sheu, H. Wu, and L.-Y. Hsu, Exploring laser-driven quantum phenomena from a time-frequency analysis perspective: A comprehensive study, Opt. Express \textbf{23}, 30459 (2015).

\bibitem{LiP}
P.-C. Li, Y.-L. Sheu, C. Laughlin, and S.-I Chu, Dynamical origin of near- and below-threshold harmonic generation of Cs in an intense mid-infrared laser field, Nat. Commun. \textbf{6}, 7178 (2015).

\end{thebibliography}
\end{document}